\documentclass[published]{JHEP3} 
\usepackage{amstext}
\usepackage{amssymb}
\usepackage{esint}
\JHEP{00(0000)000}

\JHEPspecialurl{http://jhep.sissa.it/JOURNAL/JHEP3.tar.gz}

\usepackage{epsfig,multicol,bbm}

\newcommand\fverb{\setbox\fverbbox=\hbox\bgroup\verb}
\newcommand\fverbdo{\egroup\medskip\noindent%
			\fbox{\unhbox\fverbbox}\ }
\newcommand\fverbit{\egroup\item[\fbox{\unhbox\fverbbox}]}
\newbox\fverbbox

\title{Invariance of the Hamilton-Jacobi tunneling method for black holes and FRW model}

\author{Yi-Xin Chen\\
	 Zhejiang Institute of Modern Physics, Zhejiang University,
Hangzhou, 310027, China\\
	E-mail: \email{yxchen@zimp.zju.edu.cn}}

\author{Kai-Nan Shao\\
	 Zhejiang Institute of Modern Physics, Zhejiang University,
Hangzhou, 310027, China\\
	E-mail: \email{shaokn@gmail.com}}

\received{000000} 		
\accepted{000000}		

\preprint{\hepth{000000}}	

\abstract{In this paper we revisit the topic of Hawking radiation as tunneling.
We show that the imaginary part of the action of the tunneling particle
should be reconstructed in a covariant way, as a line integral along
the classical forbidden trajectory of tunneling particles. As the
quantum tunneling phenomenon, the probability of tunneling is related
to the imaginary part of the action for the classical forbidden trajectory.
We do the calculations for massless and massive particles, in Schwarzschild
coordinate and Painlev\'e coordinate. The construction of particle action
is invariant under coordinate transformations, so this method of calculation
black hole tunneling does not have the so called ``factor 2 problem''.
As an application, we find that the temperature of Hawking temperature
of apparent horizon in a FRW universe is $T=\frac{\kappa}{2\pi}$.
Based on this result, we briefly discuss the unified first law of
apparent horizon in FRW universe.}

\keywords{Black Holes, Classical Theories of Gravity, Cosmology of Theories
beyond the SM}

\begin{document} 


\section{Introduction}

Since Hawking discovered the remarkable fact of black hole radiation\cite{Hawking1975}
in 1975, much work have done to calculate and understand this quantum
effect\cite{Hartle:1976tp}. In \cite{Kraus1995,Kraus1995a}, Kraus
and Wilczek introduced a semiclassical treatment of Hawking radiation,
by interpreting the exponent of the classical action as the modes
of the system. Subsequently, in \cite{Parikh2000} Hawking radiation
was interpreted as a tunneling phenomenon. This interpretation has
been extensively studied\cite{Chowdhury2008,Mitra2007,Pilling2008,Pilling2008a,Banerjee:2009wb,Banerjee:2009pf,Majhi:2009xh,Banerjee:2010be,Banerjee:2008sn},
and applied on various models of black holes\cite{Kar:2006iz,Zhang:2005wn,Jiang:2005ba,Kerner2008a,Jian2009}.
There are two methods to calculate the imaginary part of the action:
one is Parikh-Wilczek's radial null geodesic method\cite{Parikh2000},
and the other is the Hamilton-Jacobi method\cite{Srinivasan1999,Shankaranarayanan2001,Shankaranarayanan2002,Shankaranarayanan2003}.
Based on the Hamilton-Jacobi method, Banerjee and Majhi\cite{Banerjee2008}
developed the tunneling method beyond semiclassical approximation
to include quantum corrections. This method has been applied to calculate
quantum corrections to black hole entropy\cite{Banerjee2009a,Modak:2008tg,Banerjee2009,Zhu2009,Zhu2009a,Akbar2010}.
At first the tunneling amplitude calculated by the Hamilton-Jacobi
method differed by a factor of 2 with the standard Hawking temperature,
which is the so called ``factor of 2 problem''\cite{Chowdhury2008,Pilling2008}.
Later on, it is pointed out in \cite{Akhmedov2008,Akhmedova2008,Pilling2008a}
that by adding the temporal contribution to the imaginary part of
the action, one can resolve this problem. On the other hand, \cite{Mitra2007}
proposed a method to obtain the standard result, by introducing an
integration constant into the action, setting the incoming probability
to unity, and taking the ratio of the outgoing and incoming probabilities.
In this paper, we revisit the Hamilton-Jacobi method of calculating
black hole tunneling. Inspired by the calculation in \cite{Criscienzo},
we construct the imaginary part of the action of the tunneling particle
in an invariant way, as an integration along the classical forbidden
trajectory of the particle. By interpreting Hawking radiation as a
quantum tunneling phenomenon, the tunneling probability is related
to the imaginary part of the action of the classical forbidden trajectory\cite{landau1958quantum}.
Once again the tunneling probability is related to the Boltzmann factor
for the emission at the Hawking temperature. We perform the calculation
of black hole tunneling in Schwarzschild coordinate for massless and
massive particle, and in Painlev\'e coordinate. Since the action is
constructed invariantly, this calculation does not have the {}``factor
of 2 '' problem, and the temporal contribution has been intrinsically
included. The result shows that Hawking radiation can indeed be viewed
as a tunneling phenomenon.

In \cite{Cai:2005ra}, by assuming a temperature $T=\frac{1}{2\pi R_{H}}$
and the entropy $S=\frac{A}{4}$, where $R_{H}$ and $A$ are the
radius and area of the apparent horizon, one can introduce the first
law of thermodynamics associated with the apparent horizon, and show
that the Friedmann equation which describes the dynamics of the FRW
universe can be derived from it. There is also another proposal of
the unified first law\cite{Cai:2006rs} of the FRW model, which assigns
the temperature $T=\frac{\kappa}{2\pi}$. The tunneling method provides
a way to calculate the Hawking temperature associated with the apparent
horizon in the FRW universe. The original use of tunneling method
on FRW universe gave the result $T=\frac{1}{2\pi R_{H}}$ in \cite{Cai2009,Li:2008gf}.
However, in their calculation, the action of tunneling particle was
not constructed in the invariant way. Later on, Hayward\cite{Criscienzo}
used the invariant Hamilton-Jacobi tunneling method on dynamical black
holes, FRW model with $k=0$, gave the result $T=\frac{\kappa}{2\pi}$,
and discussed the unified first law. In dynamical black hole models,
the Hawking temperature of black hole radiation should be proportional
to surface gravity on the horizon\cite{Hayward:1997jp,Hayward:2008jq}.
The metric of FRW universe model can be viewed as a example case of
dynamical black holes. A direct thought is that their temperatures
and unified first laws should coincide. In this paper, based on the
invariant Hamilton-Jacobi method of black hole tunneling, we shall
calculate the Hawking temperature for the FRW universe with general
$k$. We get the result $T=\frac{\kappa}{2\pi}$ , and show that by
demanding the Friedmann equation to satisfy, the first law of thermodynamics
of the apparent horizon in FRW universe coincides with the unified
first law of dynamical black holes discussed in \cite{Hayward:1997jp}. Another calculation of the temperature for FRW model, see\cite{Hu:2010tx}.

The paper is organized as follows. In Section \ref{sec:The-tunneling-method},
we revisit Hamilton-Jacobi method of black hole tunneling. We show
that the imaginary part of the action of tunneling particles can be
construction in an invariant way, as a line integral along the classical
forbidden trajectory. To illustrate the method, we perform the calculation
in Schwarzschild coordinate for massless and massive particle, and
in Painlev\'e coordinate. In Section \ref{sec:Hawking-radiation-of},
we calculate the temperature associated with the apparent horizon
in FRW universe, obtain the result $T=\frac{\kappa}{2\pi}$, and then
we discuss the first law based on this temperature. Section \ref{sec:Discussions}
is for discussions. 

\section{The invariant Hamilton-Jacobi tunneling method for black holes\label{sec:The-tunneling-method}}

In the tunneling interpretation of Hawking radiation, the probability
of radiation is related to the imaginary part of the tunneling particle
via\begin{equation}
\Gamma\sim e^{-2\text{Im}I}\quad,\end{equation}
which is in turn related to the Boltzmann factor for the emission
at the Hawking temperature\begin{equation}
\Gamma\sim e^{-\frac{\omega}{T_{H}}}\quad.\end{equation}
 The semiclassical wave function of the particle is written as\begin{equation}
\phi=e^{-\frac{i}{\hbar}I}\quad.\label{eq:wavefunction}\end{equation}
The wave function satisfies the corresponding wave equation: Klein-Gorden
equation for scalar particles, Dirac equation for spin-1/2 particles,
etc. 

The action is reconstructed in an invariant way\cite{Criscienzo,diCriscienzo:2010cp},
as a line integral along the classical forbidden trajectory $\gamma$
of the tunneling particle\begin{equation}
I=\int_{\gamma}\partial_{i}I\, dx^{i}\quad.\label{eq:action}\end{equation}

Now we take the scalar particle as the example to calculate the black
hole tunneling probability.

\subsection{Schwarzschild coordinate}

The static spherical symmetric metric is written as\begin{equation}
ds^{2}=-f(r)dt^{2}+\frac{1}{f(r)}dr^{2}+r^{2}d\Omega^{2}\quad.\label{eq:metric}\end{equation}
This metric can describe a black hole whose horizon is $r_{0}$, with
$f(r_{0})=0$. 

The wave equation (\ref{eq:wavefunction}) for scalar particle satisfies
Klein-Gorden equation\begin{equation}
-\frac{\hbar^{2}}{\sqrt{-g}}\partial_{\mu}\left(g^{\mu\nu}\sqrt{-g}\partial_{\nu}\right)\phi=0\quad.\end{equation}
The leading order in $\hbar$ gives the Hamilton-Jacobi equation\begin{equation}
g^{\mu\nu}\partial_{\mu}I\partial_{\nu}I=0\quad.\label{eq:HJ_massless_sh}\end{equation}
For radial trajectory only the $(t,r)$ sector of the metric (\ref{eq:metric})
is relevant, \begin{equation}
-f(r)^{2}\left(\partial_{r}I\right)^{2}+\left(\partial_{t}I\right)^{2}=0\quad,\label{eq:HJeq_sh}\end{equation}
i.e.\begin{equation}
\partial_{r}I=\frac{1}{f(r)}\partial_{t}I\quad.\label{eq:Irt_massless_sh}\end{equation}
For outgoing modes we should take the $+$ sign. On the other hand,
the timelike Killing vector is $K=\frac{\partial}{\partial t}$, the
energy of particle is\begin{equation}
\omega=K^{i}\partial_{i}I=\partial_{t}I\quad.\label{eq:w_massless_sh}\end{equation}

The trajectory of massless particle in metric (\ref{eq:metric}) is
the null geodesics, described by $ds^{2}=0$\begin{equation}
-f(r)dt^{2}+\frac{1}{f(r)}dr^{2}=0\quad,\end{equation}
i.e.\begin{equation}
dt=\frac{1}{f(r)}dr\quad.\label{eq:traj_massless_sh}\end{equation}
Here we have chosen the $+$ sign, i.e., the outgoing mode, which
is the classical forbidden one (Classically particles are not allowed
to fall out of the black hole horizon). The curve of trajectory can
be parametrized by the variable $r$ as $\vec{\gamma}(r)=(t(r),r)$,
where $t(r)$ is determined by (\ref{eq:traj_massless_sh}). Now the
action is reconstructed as a line integral along $\gamma$\begin{eqnarray}
I & = & \int_{\gamma}dx^{i}\partial_{i}I\nonumber \\
 & = & \int_{\gamma}\left(\partial_{t}I\, dt+\partial_{r}I\, dr\right)\nonumber \\
 & = & \int\left((\partial_{t}I,\partial_{r}I)\cdot\frac{d}{dr}\vec{\gamma}(r)\right)dr\nonumber \\
 & = & \int\left(\omega\frac{1}{f(r)}dr+\frac{1}{f(r)}\omega dr\right)\nonumber \\
 & = & \int\left(\frac{2\omega}{f(r)}\right)dr\quad.\end{eqnarray}
The third line is the definition of line integral. To get the fourth
line, one can start from the second line, replace $dt$ with the trajectory
equation (\ref{eq:traj_massless_sh}) of the particle, replace $\partial_{t}I$,
$\partial_{r}I$ with the energy of the tunneling particle $\omega$
by Eqs.(\ref{eq:Irt_massless_sh}) and (\ref{eq:w_massless_sh}),
and finally remove the $\gamma$ below the the symbol of integral.
This integral is performed along the outgoing trajectory, i.e., from
inside the horizon $r_{in}$ to outside the horizon $r_{out}$. It
has a pole at the horizon $r_{0}$, where $f(r)=f(r_{0})\left(r-r_{0}\right)+O\left(r-r_{0}\right)$.
The other part of the integration is regular, so the imaginary part
of the action is\begin{equation}
\text{Im}I=\frac{2\pi\omega}{f^{\prime}(r_{0})}\quad.\end{equation}

The emission rate of the radiation is associated with the imaginary
part of the action, and it is also related with the temperature \begin{equation}
\Gamma\sim e^{-2\frac{\text{Im}I}{\hbar}}\sim e^{-\frac{\omega}{T_{H}}}\quad.\end{equation}
So, we obtain the temperature\begin{equation}
T_{H}=\frac{f^{\prime}(r_{0})\hbar}{4\pi}\quad.\end{equation}

On the other hand, the surface gravity on the horizon $r_{0}$ of
metric (\ref{eq:metric}) is\begin{equation}
\kappa_{H}^{2}=\frac{g^{\mu\nu}\partial_{\mu}\left(K^{2}\right)\partial_{\nu}\left(K^{2}\right)}{4K^{2}}\biggl|_{r=r_{0}}=\frac{g^{rr}\partial_{r}(-f)\partial_{r}(-f)}{-4f}\biggl|_{r=r_{0}}=-\frac{\left(f^{\prime}(r_{0})\right)^{2}}{4}\quad,\end{equation}
\begin{equation}
\kappa_{H}=\sqrt{-\kappa_{H}^{2}}=\frac{f^{\prime}(r_{0})}{2}\quad.\end{equation}
Now the temperature can be written as\begin{equation}
T_{H}=\frac{\hbar}{2\pi}\kappa_{H}\quad.\end{equation}
This is the standard Hawking temperature.

In order to illustrate the tunneling method mentioned above, especially
the invariant construction of the action of tunneling particle. We
next perform the calculation for a massive scalar particle. Its wave
function (\ref{eq:wavefunction}) satisfies the Klein-Gorden equation\begin{equation}
g^{\mu\nu}\partial_{\mu}I\partial_{\nu}I+m^{2}=0\quad,\end{equation}
i.e.\begin{equation}
-\frac{1}{f(r)}\left(\partial_{t}I\right)^{2}+f(r)\left(\partial_{r}I\right)^{2}+m^{2}=0\quad.\end{equation}
One should take the outgoing mode ($+$ sign)\begin{equation}
\partial_{r}I=\frac{1}{f(r)}\sqrt{-m^{2}f(r)+\left(\partial_{t}I\right)^{2}}\quad.\end{equation}

The energy of the particle is also Eq.(\ref{eq:w_massless_sh}). The
trajectory of the massive particle is the geodesics in the metric
(\ref{eq:metric}) . To simplify, the equation of geodesic motion
can be derived by the following Lagrangian\cite{Wald:1984rg}\[
L=\frac{1}{2}g_{\mu\nu}\dot{x}^{\mu}\dot{x}^{\nu}\]
where $\dot{x}$ means $\frac{dx}{d\tau}$. The Euler-Lagrange equation
$\frac{\partial L}{\partial x^{\mu}}-\frac{d}{d\tau}\left(\frac{\partial L}{\partial\dot{x}^{\mu}}\right)=0$
for $t$ is

\emph{\[
\frac{d}{d\tau}\left(f(r)\dot{t}\right)=0\qquad\Rightarrow\qquad f(r)\dot{t}=E\quad,\]
}where $E$ is a constant. The constant of the motion $L=-\frac{1}{2}$
gives\begin{equation}
\dot{r}=\sqrt{E^{2}-f(r)}\quad,\end{equation}
where we have taken the $+$ sign to get the outgoing trajectory.
Then we obtain the curve \begin{equation}
dt=\frac{E}{f(r)\sqrt{E^{2}-f(r)}}dr\quad.\end{equation}

Now the action is reconstructed as

\begin{eqnarray}
I & = & \int_{\gamma}\left(\partial_{t}I\, dt+\partial_{r}I\, dr\right)\nonumber \\
 & = & \int\left(\omega\frac{E}{f(r)\sqrt{E^{2}-f(r)}}dr+\frac{1}{f(r)}\sqrt{-m^{2}f(r)+\omega^{2}}dr\right)\nonumber \\
 & = & \int\frac{\frac{E\omega}{\sqrt{E^{2}-f(r)}}+\sqrt{\omega^{2}-m^{2}f(r)}}{f(r)}dr\quad.\end{eqnarray}
$r=r_{0}$ with $f(r_{0})=0$ is a pole of the above integration,
the imaginary part of the action is obtained by $\pi$ times the residue
of the integrand\begin{equation}
\text{Im}I=\frac{2\pi\omega}{f^{\prime}(r_{0})}\quad.\end{equation}
Once again the Hawking temperature is the standard result \begin{equation}
T_{H}=\frac{f^{\prime}(r_{0})\hbar}{4\pi}=\frac{\hbar}{2\pi}\kappa_{H}\quad.\end{equation}

\subsection{Painlev\'e coordinate}

The metric (\ref{eq:metric}) has a coordinate singularity at the
horizon $r=r_{0}$, which can be removed by transforming it to Painlev\'e
coordinate. The coordinate transformation $dt\rightarrow dt-\frac{\sqrt{1-f(r)}}{f(r)}dr$
gives \begin{equation}
ds^{2}=-f(r)dt^{2}+2\sqrt{1-f(r)}dt\, dr+dr^{2}+r^{2}d\Omega^{2}\quad.\end{equation}
The massless particle moves along the radially null geodesic described
by $ds^{2}=0$. The outgoing trajectory is\begin{equation}
dr=\frac{1+\sqrt{1-f(r)}}{f(r)}dr\quad.\end{equation}

The Hamilton-Jacobi equation (\ref{eq:HJ_massless_sh}) in this background
is\begin{equation}
-\left(\partial_{t}I\right)^{2}+2\sqrt{1-f(r)}\,\partial_{t}I\,\partial_{r}I+f(r)\left(\partial_{r}I\right)^{2}=0\quad.\end{equation}
One should choose the outgoing mode\begin{equation}
\partial_{r}I=\frac{1-\sqrt{1-f(r)}}{f(r)}\partial_{t}I\quad.\end{equation}
The energy of the particle is still $\omega=\partial_{t}I$, as in
(\ref{eq:w_massless_sh}).

The action of the particle is reconstructed as\begin{eqnarray}
I & = & \int_{\gamma}\left(\partial_{t}I\, dt+\partial_{r}I\, dr\right)\nonumber \\
 & = & \int\left(\omega\frac{1+\sqrt{1-f(r)}}{f(r)}dr+\omega\frac{1-\sqrt{1-f(r)}}{f(r)}dr\right)\nonumber \\
 & = & \int\left(\frac{2\omega}{f(r)}\right)dr\quad.\end{eqnarray}
The same result as the massless particle in the Schwarzschild coordinate.
This is a natural result of the invariant construction of the particle
action.

\subsection{Discussion and comments on the factor 2 problem}

In the original calculation of black hole tunneling \cite{Parikh2000},
the action of particle $\text{Im}I$ is constructed as $\text{Im}\int pdr$
in the Painlev\'e coordinate, and $\Gamma\sim e^{-2\text{Im}\int pdr}$.
It is pointed out in \cite{Chowdhury2008} that the quantity $\Gamma\sim e^{-2\text{Im}\int pdr}$
is not canonically invariant, and the answer would be different in
different canonical frame. He proposed that one should use the canonically
invariant formula $\Gamma\sim e^{-Im\oint pdr}$. However, using this
canonically invariant formula in Schwarzschild coordinate, the Hawking
temperature obtained is twice as the original one \cite{Akhmedov2006}.
This is the so called ``factor of 2 problem''\cite{Pilling2008}.
Later on, this problem was resolved by adding the temporal contribution
to the action\cite{Akhmedov2008,Akhmedova2008,Pilling2008a,Zhu:2009wa}. The
calculation in Painlev\'e coordinate is correct using $\Gamma\sim e^{-2\text{Im}\int pdr}$,
as presented in \cite{Pilling2008a}. Other solutions of this problem
is to introduce an integration constant into the action\cite{Mitra2007,Stotyn:2008qu},
or consider the thermal balance and take the tunneling rate as the
ratio between the emission and absorption probabilities\cite{Hu:2009me}.

In our paper, we construct the action as (\ref{eq:action}), inspired
by the calculation in \cite{Criscienzo}. As mentioned above, the
action is a scalar quantity and should be canonically invariant. Eq.(\ref{eq:action})
is invariant, and gives the correct temperature. In fact, temporal
contribution has intrinsically been included in this construction
of action. Additionally, we illustrate that the integration in (\ref{eq:action})
is in fact a line integral, and should be performed along the classical
forbidden trajectory (here it is the trajectory from inside to outside
of black hole horizon) of the test particle. This coincides with the
philosophy of quantum tunneling\cite{landau1958quantum}, which tells
that the propagating rate of quantum tunneling is related to the imaginary
part of the action along classical forbidden trajectory.

\section{Hawking radiation of apparent horizon in a FRW universe and unified
first law\label{sec:Hawking-radiation-of}}

The tunneling method can also be applied to analyze the radiation
of dynamical black holes\cite{Hayward:2008jq,Criscienzo,diCriscienzo:2010cp}
and FRW universe model\cite{Cai2009,Li:2008gf}. The calculation of
Hawking radiation of the apparent horizon in FRW method using scalar
particles tunneling\cite{Cai2009} and fermions tunneling\cite{Li:2008gf}
gives the temperature $T=\frac{1}{2\pi R_{H}}$. However, the action
of the tunneling particles through the apparent horizon are not constructed
in the invariant way. Later on, Hayward\cite{Hayward:2008jq} calculated
the temperature of dynamical black holes, and gave the result $T=\frac{\kappa}{2\pi}$.
The temperature of dynamical black hole and FRW universe with $k=0$
were also calculated by the invariant Hamilton-Jacobi method \cite{Criscienzo}.
Using the Hawking temperature, on can discuss the first law of thermodynamics
associated with the apparent horizon of dynamical black holes and
FRW model. In fact, the FRW metric can be viewed as a certain kind
of dynamic black hole. In this section we shall apply the tunneling
method illustrated in the above section to the FRW model with general
$k$, obtain its temperature, and discuss the first law on its apparent
horizon.

The homogeneous and isotropic universe model is described by the $(3+1)$
dimensional FRW metric\begin{equation}
ds^{2}=-dt^{2}+\frac{a(t)^{2}}{1-kr^{2}}dr^{2}+a(t)^{2}r^{2}d\Omega^{2}\quad,\label{eq:FRW1}\end{equation}
where the spatial curvature constant $k=+1$, $0$ and $-1$, corresponding
to a closed, flat and open universe, respectively. Defining $R=a(t)r$,
$H(t)=\frac{\dot{a}}{a}$ and $R_{A}(t):=\frac{1}{\sqrt{H^{2}+\frac{k}{a(t)^{2}}}}$.
, the metric (\ref{eq:FRW1}) can be rewritten as\begin{equation}
ds^{2}=-\frac{1-\frac{R^{2}}{R_{A}^{2}}}{1-\frac{kR^{2}}{a^{2}}}dt^{2}-\frac{2HR}{1-\frac{kR^{2}}{a^{2}}}dt\, dR+\frac{1}{1-\frac{kR^{2}}{a^{2}}}dR^{2}+R^{2}d\Omega^{2}\quad.\label{eq:FRW2}\end{equation}
Denote the $(t,R)$ sector of the above metric by $ds^{2}=\gamma_{ij}dx^{i}dx^{j}$,
$(i,j=\{0,1\})$. Now we can calculate the quantities mentioned in
\cite{Criscienzo}.

The dynamical apparent horizons $\mathcal{H}$ is a marginally trapped
surface with vanishing expansion, determined by\begin{equation}
\chi(x)\biggl|_{H}=\gamma^{ij}\partial_{i}R\partial_{j}R\biggl|_{H}=0\end{equation}
i.e.\begin{equation}
R_{H}==\frac{1}{\sqrt{H^{2}+\frac{k}{a(t)^{2}}}}=R_{A}(t)\quad.\label{eq:FRW_horizon}\end{equation}

The Misner-Sharp mass, on the horizon, is\begin{equation}
m_{H}=\frac{1}{2}R(1-\chi)\biggl|_{H}=\frac{R_{H}}{2}\quad.\end{equation}

In spherical symmetric case, it is possible to introduce the Kodama
vector field\begin{equation}
K^{i}=\frac{1}{\sqrt{-\gamma}}\varepsilon^{ij}\partial_{j}R\,,\qquad K^{\theta}=0=K^{\varphi}\quad,\end{equation}
where $\epsilon_{ij}=\frac{1}{\sqrt{1-kR^{2}/a^{2}}}(dt)_{i}\wedge(dR)_{j}$.
Here $K^{i}=\left(\sqrt{1-k\frac{R^{2}}{a(t)^{2}}},0\right)$. The
Kodama vector gives a preferred flow of time and is a dynamic analogue
of a stationary Killing vector. By using the Kodama vector, the Misner-Sharp
mass can be defined as a conserved quantity\cite{Hayward:1994bu}.
The Kodama vector is timelike, null and spacelike as $R<R_{H}$, $R=R_{H}$
and $R>R_{H}$. One can also introduce a dynamical surface gravity\cite{Hayward:1997jp,Hayward:2008jq,Criscienzo}
$\kappa$ defined by\begin{equation}
\kappa=\frac{1}{2}\Box_{\gamma}R\quad,\end{equation}
and satisfies similar Killing identity on the horizon \begin{equation}
K^{a}\nabla_{[b}K_{a]}\cong\pm\kappa K_{b}\quad.\end{equation}
The surface gravity associated with the apparent horizon is

\begin{eqnarray}
\kappa_{H} & = & \frac{1}{2\sqrt{-\gamma}}\partial_{i}\left(\sqrt{-\gamma}\gamma^{ij}\partial_{j}R(x)\right)\biggl|_{H}\nonumber \\
 & = & -\frac{R\left(k+a^{2}\left(2H^{2}+\dot{H}\right)\right)}{2a^{2}}\biggl|_{H}\nonumber \\
 & = & -\frac{R}{2}\left(\frac{k}{a^{2}}+2H^{2}+\dot{H}\right)\biggl|_{H}\nonumber \\
 & = & -\frac{R_{H}}{2}\left(\dot{H}+2H^{2}+\frac{k}{a^{2}}\right)\nonumber \\
 & = & -\frac{a^{2}\left(2H^{2}+\dot{H}\right)+k}{2\sqrt{a^{2}H^{2}+k}}\quad.\end{eqnarray}

In \cite{Cai:2005ra,Cai:2006rs,Gong:2007md}, by assuming a temperature
to the apparent horizon, one can discuss the first law of thermodynamics
in the FRW universe. Later on, \cite{Cai2009,Li:2008gf} proposed
that one can derive the temperature of the apparent horizon using
the tunneling method. The energy of particle tunneling through the
apparent horizon is defined using the Kodama vector\begin{equation}
\omega=-K^{i}\partial_{i}I=-\sqrt{1-k\frac{R^{2}}{a^{2}}}\partial_{t}I\quad.\label{eq:w_FRW}\end{equation}

For a massless particle, the Hamilton-Jacobi equation (\ref{eq:HJ_massless_sh})
is\begin{equation}
-(\partial_{t}I)^{2}-2RH(\partial_{R}I)(\partial_{t}I)+\left(1-\frac{kR^{2}}{a^{2}}-R^{2}H^{2}\right)(\partial_{R}I)^{2}=0\quad.\end{equation}
For the outgoing mode we take the plus sign\begin{equation}
\partial_{t}I=\left(\sqrt{1-k\frac{R^{2}}{a^{2}}}-RH\right)(\partial_{R}I)\quad.\label{eq:It_FRW}\end{equation}

The trajectory for the massless particle is the radial null geodesics,
determined by \begin{equation}
0=ds^{2}=-\frac{1-\frac{R^{2}}{R_{A}^{2}}}{1-\frac{kR^{2}}{a^{2}}}dt^{2}-\frac{2HR}{1-\frac{kR^{2}}{a^{2}}}dt\, dR+\frac{1}{1-\frac{kR^{2}}{a^{2}}}dR^{2}\quad.\end{equation}
For the outgoing particle, the curve is\begin{equation}
dR=\left(\sqrt{1-k\frac{R^{2}}{a^{2}}}+RH\right)dt\quad.\label{eq:traj_FRW}\end{equation}
The curve can be parametrized by the variable $R$ as $\vec{\gamma}(R)=\left(t(R),R\right)$,
where the function $t(R)$ is determined by (\ref{eq:traj_FRW}). 

The action of the particle tunneling through the horizon can be reconstructed
as a line integral along the incoming trajectory\begin{equation}
I=\int_{\gamma}dx^{i}\partial_{i}I=\int_{\gamma}\left(dt\partial_{t}I+dR\partial_{R}I\right)\quad.\end{equation}
Combining Eqs. (\ref{eq:w_FRW})(\ref{eq:It_FRW})(\ref{eq:traj_FRW}),
the action can be calculated as\begin{eqnarray}
I & = & \int\left(-\frac{\text{\ensuremath{\omega}}}{\sqrt{1-k\frac{R^{2}}{a^{2}}}}\frac{1}{\sqrt{1-k\frac{R^{2}}{a^{2}}}+RH}-\frac{\text{\ensuremath{\omega}}}{\sqrt{1-k\frac{R^{2}}{a^{2}}}}\frac{1}{\sqrt{1-k\frac{R^{2}}{a^{2}}}-RH}\right)dR\nonumber \\
 & = & -\int\left(\frac{2\omega}{1-H^{2}R^{2}-\frac{kR^{2}}{a^{2}}}\right)dR\end{eqnarray}
Note that we should express the above integrand as a function of $R$,
say, $H(t(R))$ and $a(t(R))$, where the function $t(R)$ is is determined
by (\ref{eq:traj_FRW}). This integral has a pole at the horizon $R=R_{H}$,
where\[
1-H^{2}R^{2}-\frac{kR^{2}}{a^{2}}=\left(1-\frac{kR_{H}^{2}}{a^{2}}-R_{H}^{2}H^{2}\right)+\left(-\frac{a^{2}\left(2H^{2}+\dot{H}\right)+k}{\sqrt{a^{2}H^{2}+k}}\right)(R-R_{H})+O(R-R_{H})\quad.\]
The first term on the right hand side equals zero, due to the definition
of the horizon (\ref{eq:FRW_horizon}), and the coefficient of the
second term gives $-2\left|\kappa_{H}\right|$. The imaginary part
of the action is then\begin{equation}
\text{Im}I=\frac{\pi\omega}{\left|\kappa_{H}\right|}\quad.\end{equation}

Again, the emission rate of the radiation is associated with the imaginary
part of the action, and it is also related with the temperature \[
\Gamma\sim e^{-2\frac{\text{Im}I}{\hbar}}\sim e^{-\frac{\omega}{T_{H}}}\quad.\]
So, we obtain the temperature for Hawking radiation of the apparent
horizon in FRW universe\begin{equation}
T_{H}=\frac{\left|\kappa_{H}\right|}{2\pi}\quad.\label{eq:T_FRW}\end{equation}

Using the expression (\ref{eq:T_FRW}) of the temperature, we can
discuss the first law of thermodynamics associated with the apparent
horizon. According to \cite{Hayward:1997jp,Cai:2006rs}, the unified
first law is written as \begin{equation}
dm_{H}=T_{H}dS+WdV_{H}\quad,\label{eq:FRW_1stlaw}\end{equation}
where $S=\frac{A_{H}}{4}$, a quarter of the horizon area, and $WdV$
is the work term. 

In the FRW universe, the energy-momentum tensor has the form of perfect
fluid\begin{equation}
T_{ab}=(\rho+p)U_{a}U_{b}+pg_{ab}\quad,\end{equation}
where $U^{a}$ denotes the four-velocity of the fluid, $\rho$, $p$
are the energy density and pressure, respectively. In the FRW metric
(\ref{eq:FRW1}), the components of $T_{ab}$ are\begin{equation}
T_{00}=\rho,\qquad T_{ij}=pg_{ij}\quad.\end{equation}
Write them out explicitly, in the $(t,r)$ sector \begin{equation}
T_{ab}^{(2)}=\left(\rho,\frac{p\, a^{2}}{1-kr^{2}}\right)\quad,\qquad T_{\,\,\, a}^{(2)\, b}=\text{diag}\left(-\rho,p\right)\quad.\end{equation}
The work term is given by \begin{equation}
W=-\text{trace}T^{(2)}=\frac{1}{2}\left(\rho-p\right)\quad.\end{equation}

Substitute the expressions\[
m_{H}=\frac{R_{H}}{2}\quad,\qquad A_{H}=4\pi R_{H}^{2}\]
\[
V_{H}=\frac{4}{3}\pi R_{H}^{2}\quad,\qquad\kappa_{H}=-\frac{R_{H}}{2}\left(\dot{H}+2H^{2}+\frac{k}{a^{2}}\right)\]
and the temperature $T_{H}=\frac{\left|\kappa_{H}\right|}{2\pi}$,
comparing the coefficients of $dR_{H}$ on both sides, and demanding
that the Friedmann equations\begin{equation}
H^{2}+\frac{k}{a^{2}}=\frac{8\pi}{3}\rho\end{equation}
\begin{equation}
\dot{H}-\frac{k}{a^{2}}=-4\pi(\rho+p)\end{equation}
one can check that the unified first law (\ref{eq:FRW_1stlaw}) in
deed holds.

Thus, we have obtained the Hawking temperature of the apparent horizon
in the FRW universe, and check that this temperature can indeed be
consistent with a kind of unified first law of thermodynamics. This
is the application of the tunneling method discussed in the previous
section in the FRW universe model.

\section{Discussions\label{sec:Discussions}}

In this article, we revisited the topic of interpreting Hawking radiation
of black holes as an effect of quantum tunneling. We discuss the Hamilton-Jacobi
method of black hole tunneling. The main viewpoint of this article
is that the action should be constructed as a line integral along
the classically forbidden trajectory of the tunneling particle. This
invariant construction is mostly inspired by the discussion in \cite{Criscienzo}.
As an example, we illustrate the calculation for massless particles
in Schwarzschild coordinate. Additionally, in order to support this
method, we perform the calculation for a massive particle, and for
a massless particle in Painlev\'e coordinate. These calculations all
give the result of standard Hawking temperature. Since the propagating
rate of quantum tunneling is related to the imaginary part of the
action along classical forbidden trajectory, the results here support
the interpretation of Hawking radiation as a quantum tunneling phenomenon.
On the other hand, since the construction of the action is invariant,
there is not the ``factor of 2 problem'' here. 

The thermodynamics of the FRW model has been extensively discussed.
As an application, we calculate this temperature using the Hamilton-Jacobi
tunneling method discussed in this article. We obtain the result $T=\frac{\left|\kappa_{H}\right|}{2\pi}$,
and show that the unified first law based on this expression of temperature
can be consistent with the Friedmann equations which describe the
dynamical properties of the FRW model. This result is different with
the temperature $T=\frac{1}{2R_{H}}$ in\cite{Cai2009,Li:2008gf}.
In fact, in their calculation the actions of tunneling particles are
not constructed in an invariant way. Maybe the result $T=\frac{1}{2R_{H}}$
can be viewed as a certain kind of approximation on the apparent horizon.
Another point is that in \cite{Cai2009,Li:2008gf} they chose the
in coming mode for the tunneling particles, while in \cite{Criscienzo}
and our article the choice is the outgoing mode. This point still
needs further investigation.

\begin{acknowledgments}

We thank C.Cao, Q.J.Cao, Y.J.Du, J.L.Li and Q.Ma for useful discussions.
Chen would like to thank the organizer and the participants of the
advanced workshop, \textquotedblleft{}Dark Energy and Fundamental
Theory\textquotedblright{} supported by the Special Fund for Theoretical
Physics from the National Natural Science Foundation of China with
grant No. 10947203, for stimulating discussions and comments. The
research is supported by the NNSF of China Grant No. 90503009, No.
10775116, 973 Program Grant No. 2005CB724508, and in part by the Project
of Knowledge Innovation Program (PKIP) of Chinese Academy of Sciences,
Grant No. KJCX2.YW.W10. 

\end{acknowledgments}

\bibliographystyle{JHEP}
\bibliography{InvarianceTunneling}

\end{document}